# Controlling chaos in wave-particle interactions

M. C. de Sousa[1], I. L. Caldas[1], F. B. Rizzato[2], R. Pakter[2], F. M. Steffens[1]
[1]*Instituto de Física, Universidade de São Paulo, São Paulo, SP, Brazil*
[2]*Instituto de Física, Universidade Federal do Rio Grande do Sul, Porto Alegre, RS, Brazil*

We analyze the behavior of a relativistic particle moving under the influence of a uniform magnetic field and a stationary electrostatic wave. We work with a set of pulsed waves that allows us to obtain an exact map for the system. We also use a method of control for near-integrable Hamiltonians that consists in the addition of a small and simple control term to the system. This control term creates invariant tori in phase space that prevent chaos from spreading to large regions and make the controlled dynamics more regular. We show numerically that the control term just slightly modifies the system but is able to drastically reduce chaos with a low additional cost of energy. Moreover, we discuss how the control of chaos and the consequent recovery of regular trajectories in phase space are useful to improve regular particle acceleration.

## I. INTRODUCTION

Wave-particle interaction has always attracted a great deal of attention since it can be found in many fields, such as in particle accelerators [1, 2], free electron lasers [2], cyclotron autoresonance [2], astrophysical systems [2], in the description of the microscopic dynamics of plasma [2] and in current drives in fusion devices [3]. Besides, wave-particle interaction presents a wide range of applications as an efficient way for particle heating [2, 4--6] and particle acceleration [2, 6--8].

Wave-particle interaction is basically a nonlinear process [2, 9] that can present regular and chaotic trajectories in its phase space [10]. Periodic or quasiperiodic trajectories are useful for coherent acceleration [11], while chaotic trajectories are responsible for particle heating [4]. The prevalence of one or another pattern depends mainly on the amplitude of the perturbation applied to the system.

As a general rule, the system becomes more and more chaotic as the amplitude of the perturbation increases. But for this kind of interaction, the behavior of the particles usually becomes chaotic even for small wave amplitudes. This characteristic makes it important to control chaos in the system, so one can have regular behavior for larger values of the amplitude of the wave.

Control of chaos is a key challenge in many areas of physics [12--21] and several methods have been developed with the purpose of controlling a few specific trajectories in phase space [21--26]. But for systems where we have wave-particle interaction, these methods are hopeless because of the great number of trajectories to deal with simultaneously. Some years ago, a method based on perturbation theory and Lie algebra was developed in Ref. [27] for conservative systems that can be described by an integrable Hamiltonian plus a small perturbation. Instead of controlling some chosen trajectories, this method consists in creating invariant tori in phase space that prevent chaos from occurring.

However, the control term used to create the invariant tori in phase space must fit some conditions as follows. For energetical purposes, the control term is required to be much smaller than the perturbation originally applied to the system, i.e., it should just slightly modify the system but be able to reduce chaos in phase space. It is also required that the control term must be as simple as possible in order to be implemented in experiments such as in Ref. [28] where a control term was used to increase the kinetic coherence of an electron beam in a traveling wave tube.

Besides the experiment mentioned above, the method of control developed in Ref. [27] has been used mainly as a way to control chaotic transport in Hamiltonian systems, such as in magnetized plasmas [19, 29, 30], fusion devices [19] and turbulent electric fields [31]. Indeed, to control chaos in the system, this method creates invariant tori in the whole phase space preventing chaotic transport from taking place. However, the addition of invariant tori to the system also restores many of the original regular trajectories that otherwise would have been destroyed by chaos. As periodic and quasiperiodic trajectories are responsible for coherent acceleration, the recovery of these trajectories is very important if one is interested in particle acceleration.

In this paper, we show how the control of chaos is useful to improve regular particle acceleration. To do so, we analyze the behavior of a relativistic particle moving under the combined action of a uniform magnetic field and a stationary electrostatic wave. External magnetic fields are usually present in systems where we have wave-particle interaction, such as in astrophysical systems. Besides, magnetic fields are useful in experiments because they restrict the movement of the particle confining it to a limited region of space.

In our model, we consider the electrostatic wave as a series of pulses that perturbs the system periodically in time. Such a kicked model allows us to integrate the Hamiltonian analytically between two consecutive pulses and it is possible to obtain a Poincaré map that describes the time evolution of the system.

To control chaos in the system, we use the method of control developed in Ref. [27] and presented in Ref. [28--30], which consists in the addition of a small perturbation to the Hamiltonian. For the system under study, we show that this perturbation is simply a second stationary electrostatic wave with wave amplitude much smaller than the amplitude of the original electrostatic wave. The second wave added to the system does not change its main structures but makes the dynamics regular for larger values of the amplitude of the original electrostatic wave.

By regularizing the system and recovering periodic and quasiperiodic trajectories, the control of chaos also improves the coherent acceleration that the particle experiences in these trajectories. It means that the initial energy of the particle can be lower in the controlled dynamics than it should be in the system without control and even so its final energy will be higher than it is in the original system.

The paper is organized as follows: in Section II we describe our system of wave-particle interaction; in Section III we briefly present the method of control that we use and we calculate the control term for our system; in Section IV we present the analytical and numerical results obtained, including the control of chaos in our system, the improvement in the acceleration of the particle, and a qualitative analysis about the action of the second electrostatic wave added to the Hamiltonian and how it is able to control chaos in the system; in Section V we present our conclusions.

## II. DESCRIPTION OF THE SYSTEM

In this paper, we analyze a beam of charged particles interacting with a magnetic field and an electrostatic wave. We consider a very low density beam to be sure that the beam does not induce any wave growth. Moreover, the particles of a low density beam can be considered as test particles that do not interact with each other. The main effect in this case is the individual interaction of each particle with the magnetic field and the electrostatic wave as it will be considered next.

Following Ref. [32], suppose a relativistic particle with charge $q$, mass $m$, and canonical momentum $\mathbf{p}$ moving under the combined action of a uniform magnetic field $\mathbf{B} = B_0 \hat{\mathbf{z}}$ and a stationary electrostatic wave of wave vector $k$, period $T$, and amplitude $\varepsilon/2$ lying along the $x$ axis. The transversal dynamics of this system is described by the Hamiltonian

$$H = \sqrt{m^2 c^4 + c^2 p_x^2 + c^2 (p_y + qB_0 x)^2} + \frac{\varepsilon}{2}\cos(kx) \sum_{n=-\infty}^{+\infty} \delta(t-nT), \quad (1)$$

where $c$ is the speed of light and we chose to work with a pulsed system that is represented by the periodic collection of delta functions.

As Hamiltonian (1) is not a function of the $y$ variable, we see that $p_y$ is a constant of motion and for simplicity we will assume, with no loss of generality, $p_y = 0$. We remark that although $p_y$ is conserved and we are taking it to be zero, $dy/dt$ is not zero and the transversal motion of the particle is not one dimensional.

Working with the dimensionless quantities $qB_0(t,T)/m \to (t,T)$, $qB_0 x/mc \to x$, $mck/qB_0 \to k$, $p_x/mc \to p_x$ and $(qB_0/m^2c^2)\varepsilon \to \varepsilon$, the dimensionless Hamiltonian $H/mc^2 \to H$ that describes the system is given by

$$H = \sqrt{1 + p_x^2 + x^2} + \frac{\varepsilon}{2}\cos(kx) \sum_{n=-\infty}^{+\infty} \delta(t-nT). \quad (2)$$

From this expression, we observe that between two consecutive pulses, Hamiltonian (2) becomes integrable and time independent. Thus, it is possible to write Hamiltonian (2) in terms of its action-angle variables through the canonical transformation $x = \sqrt{2I}\sin\theta$ and $p = \sqrt{2I}\cos\theta$. In the variables $(I, \theta)$, the dynamics of the system is described by

$$H = \sqrt{1 + 2I} + \frac{\varepsilon}{2}\cos\left(k\sqrt{2I}\sin\theta\right) \sum_{n=-\infty}^{+\infty} \delta(t-nT). \quad (3)$$

At this point we must adopt a strategy to integrate the system and to obtain a map that describes its time evolution:

1-) First, we observe that the second term in Hamiltonian (2) is only a function of the $x$ variable and, therefore, just the momentum $p_x$ experiences an abrupt change in its value when $t = nT$. We calculate the change in the variables $x$ and $p_x$ across kick $n$ as $\Delta x = 0$ and $\Delta p_x = (\varepsilon k/2)\sin(kx_n)$, where $x_n$ and $p_{x,n}$ are the values of $x$ and $p_x$ immediately before kick $n$.

2-) The values of $x$ and $p_x$ immediately after kick $n$ will be given by $x_n^+ = x_n + \Delta x$ and $p_{x,n}^+ = p_{x,n} + \Delta p_x$. We also calculate the values of $I$ and $\theta$ immediately after kick $n$ as $I_n^+ = 0.5[(x_n^+)^2 + (p_{x,n}^+)^2]$ and $\theta_n^+ = \arctan(x_n^+/p_{x,n}^+)$.

3-) Between two consecutive kicks, Hamiltonian (3) is integrable and depends only on the action variable $I$, which makes it possible to calculate exactly the changes in the variables $I$ and $\theta$ between kicks $n$ and $n+1$: $\Delta I = 0$ and $\Delta \theta = T/\sqrt{1+2I_{n+1}}$.

4-) Therefore, the values of $I$ and $\theta$ immediately before kick $n+1$ will be given by $I_{n+1} = I_n^+ + \Delta I$ and $\theta_{n+1} = \theta_n^+ + \Delta\theta$.

Finally, the map that describes the time evolution of the system in variables $I$ and $\theta$ is given by Eq. (4)

$$I_{n+1} = \frac{1}{2}\left\{2I_n \sin^2\theta_n + \left[\sqrt{2I_n}\cos\theta_n + \frac{1}{2}\varepsilon k \sin\left(k\sqrt{2I_n}\sin\theta_n\right)\right]^2\right\},$$

$$\theta_{n+1} = \arctan\left(\frac{2\sqrt{2I_n}\sin\theta_n}{2\sqrt{2I_n}\cos\theta_n + \varepsilon k \sin\left(k\sqrt{2I_n}\sin\theta_n\right)}\right) + \frac{T}{\sqrt{1+2I_{n+1}}}. \quad (4)$$

The obtained map is symplectic and it has the noticeable feature of being completely explicit ($I_{n+1}$ in the second equation can be written explicitly as a function of $I_n$ and $\theta_n$ if we use the first equation of the map). Besides, map (4) has a strong nonlinear dependence on the wave amplitude $\varepsilon$.

We point out that map (4) could have been written with the same degree of complexity in the variables $(x, p_x)$, but the use of action-angle variables is more convenient since action is conserved in the absence of perturbation. Besides, it is simpler to use action-angle variables when calculating the control term for the system as will be shown in the next section.

### III. METHOD OF CONTROL

In this section, we briefly present a method of control based on Lie algebra that was proposed in Ref. [27] for conservative systems that can be described by a near-integrable Hamiltonian of the form $H = H_0 + \varepsilon V$, where $H_0$ is integrable and $\varepsilon V$ is a small perturbation with $\varepsilon \ll 1$. For $\varepsilon = 0$, the dynamics is integrable and the phase space presents just invariant tori. As $\varepsilon$ grows from zero, the trajectories are altered and the system becomes more and more chaotic.

Our purpose is to find a control term $f(\varepsilon V)$ such that the controlled Hamiltonian $H_f = H_0 + \varepsilon V + f(\varepsilon V)$ is integrable or presents a more regular behavior than the original one. In this sense, $f = -\varepsilon V$ is an obvious solution because, with this control term, the new Hamiltonian would be integrable. However, $f = -\varepsilon V$ is not suitable since it is on the same order of the original perturbation. For energetical purposes, the control term is required to be much smaller than $\varepsilon V$, for instance, if $\varepsilon V$ is on the order of $\varepsilon$, then $f(\varepsilon V)$ should be on the order of $\varepsilon^2$. Thus, what we look for is a control term that just slightly modifies the system but is able to reduce chaos in phase space.

The control term can be either a global one and acts in the entire phase space [28–30] or a local one that acts only in a specific region of phase space [19, 31, 33]. However, in this paper, we consider only the case where $f$ is a global control term. We start by taking $A$ as the Lie algebra of real-valued functions of class $\mathbb{C}^\infty$ defined in phase space. We consider an element $H \in A$ that we call a Hamiltonian. The linear operator associated with $H$ will be $\{H\}$ and it acts on $A$ such that

$$\{H\}H' = \{H, H'\}$$

for any $H' \in A$, where $\{\,,\,\}$ denotes the Poisson bracket.

Let $H_0 \in A$ be an integrable Hamiltonian written as a function of the action-angle variables $(\mathbf{I}, \boldsymbol{\theta}) \in B \times T^n$, where B is a domain of $\mathbb{R}^n$, $T^n$ is the n-dimensional torus and $n$ is the number of degrees of freedom. Using action-angle variables, the Poisson bracket between two Hamiltonians is given by

$$\{H, H'\} = \frac{\partial H}{\partial \mathbf{I}} \frac{\partial H'}{\partial \boldsymbol{\theta}} - \frac{\partial H}{\partial \boldsymbol{\theta}} \frac{\partial H'}{\partial \mathbf{I}}.$$

Consider now an element $V(\mathbf{I}, \boldsymbol{\theta}) \in A$ expanded in Fourier series as

$$V(\mathbf{I}, \boldsymbol{\theta}) = \sum_{\mathbf{k} \in \mathbb{Z}^n} V_k(\mathbf{I}) e^{i \mathbf{k} \cdot \boldsymbol{\theta}}$$

and the action of the operator $\{H_0\}$ on $V$ as

$$\{H_0\} V(\mathbf{I}, \boldsymbol{\theta}) = \sum_{\mathbf{k} \in \mathbb{Z}^n} i \boldsymbol{\omega}(\mathbf{I}) \cdot \mathbf{k}\, V_k(\mathbf{I}) e^{i \mathbf{k} \cdot \boldsymbol{\theta}},$$

where

$$\boldsymbol{\omega}(\mathbf{I}) = \frac{\partial H_0}{\partial \mathbf{I}}$$

denotes the frequency vector.

We define a pseudoinverse of $\{H_0\}$ as a linear operator $\Gamma$ on $A$

$$\{H_0\}^2 \Gamma = \{H_0\}.$$

We choose the operator $\Gamma$ such that its action on $V$ is given by

$$\Gamma V(\mathbf{I}, \boldsymbol{\theta}) = \sum_{\substack{\mathbf{k} \in \mathbb{Z}^n \\ \boldsymbol{\omega}(\mathbf{I}) \cdot \mathbf{k} \neq 0}} \frac{V_k(\mathbf{I})}{i \boldsymbol{\omega}(\mathbf{I}) \cdot \mathbf{k}} e^{i \mathbf{k} \cdot \boldsymbol{\theta}}$$

and we note that this choice of $\Gamma$ commutes with $\{H_0\}$.

We also build two other operators, $N$ and $R$, for which we have $RV$ as the resonant part of $V$ and $NV$ as its nonresonant part as follows

$$RV(\mathbf{I}, \boldsymbol{\theta}) = \sum_{\substack{\mathbf{k} \in \mathbb{Z}^n \\ \boldsymbol{\omega}(\mathbf{I}) \cdot \mathbf{k} = 0}} V_k(\mathbf{I}) e^{i \mathbf{k} \cdot \boldsymbol{\theta}},$$

$$NV(\mathbf{I}, \boldsymbol{\theta}) = \sum_{\substack{\mathbf{k} \in \mathbb{Z}^n \\ \boldsymbol{\omega}(\mathbf{I}) \cdot \mathbf{k} \neq 0}} V_k(\mathbf{I}) e^{i \mathbf{k} \cdot \boldsymbol{\theta}}.$$

Finally, the function $f(V)$ that makes the controlled system more regular than the original one is defined as

$$f(V) = \sum_{n=1}^{\infty} \frac{(-1)^n}{(n+1)!} \{\Gamma V\}^n (nR + 1) V \qquad (5)$$

and it can be proved that expression (5) is really a control term [27]. From expression (5), we notice that if $V$ is on the order of $\varepsilon$, $f(V)$ has a dominant term on the order of $\varepsilon^2$, thus much smaller than $V$ since $\varepsilon \ll 1$.

Now that we have defined all the operators, we are able to apply them to our system. But to do so, it is necessary to map the time dependent Hamiltonian given in (3) into an autonomous Hamiltonian with 2 degrees of freedom. This is performed by considering that $t \mod 2\pi$ is an additional angle variable and that $E$ is its conjugated action. Then, the autonomous Hamiltonian with 2 degrees of freedom will be

$$H(I, E, \theta, t) = E + \sqrt{1 + 2I}$$
$$+ \frac{\varepsilon}{2} \cos\left(k\sqrt{2I} \sin\theta\right) \sum_{n=-\infty}^{+\infty} \delta(t - nT). \qquad (6)$$

Besides, we will write the periodic collection of delta functions of Hamiltonian (6) as a Fourier series in order to properly compute all the Poisson brackets involved in the control theory. The resulting Hamiltonian is given by expression (7)

$$H(I, E, \theta, t) = H_0(I, E) + \varepsilon V(I, \theta, t)$$
$$H(I, E, \theta, t) = E + \sqrt{1+2I}$$
$$+ \frac{\varepsilon}{2T}\cos\left(k\sqrt{2I}\sin\theta\right)\sum_{n=-\infty}^{+\infty}\cos\left(\frac{2\pi n t}{T}\right). \quad (7)$$

The actions of the operators $\{H_0\}$, $\Gamma$, $R$ and $N$ on the perturbing Hamiltonian $\varepsilon V$ given in (7) are

$$\{H_0\}(\varepsilon V) = -\frac{2\pi\varepsilon}{2T^2}\cos\left(k\sqrt{2I}\sin\theta\right)\sum_{n=-\infty}^{+\infty}n\sin\left(\frac{2\pi n t}{T}\right)$$
$$- \frac{\varepsilon k\sqrt{2I}\cos\theta}{2T\sqrt{1+2I}}\sin\left(k\sqrt{2I}\sin\theta\right)\sum_{n=-\infty}^{+\infty}\cos\left(\frac{2\pi n t}{T}\right), \quad (8)$$

$$\Gamma(\varepsilon V) = -\frac{\varepsilon}{4\pi}\cos\left(k\sqrt{2I}\sin\theta\right)\sum_{\substack{n=-\infty\\n\neq 0}}^{+\infty}\frac{1}{n}\sin\left(\frac{2\pi n t}{T}\right)$$
$$- \frac{\varepsilon\sqrt{1+2I}}{2Tk\sqrt{2I}\cos\theta}\sin\left(k\sqrt{2I}\sin\theta\right)\sum_{n=-\infty}^{+\infty}\cos\left(\frac{2\pi n t}{T}\right)$$

($I \neq 0$ and $\theta \neq \pi/2, 3\pi/2$ for $0\leq \theta < 2\pi$), (9)

$$R(\varepsilon V) = 0, \quad (10)$$

$$N(\varepsilon V) = \frac{\varepsilon}{2T}\cos\left(k\sqrt{2I}\sin\theta\right)\sum_{n=-\infty}^{+\infty}\cos\left(\frac{2\pi n t}{T}\right). \quad (11)$$

As Hamiltonian (7) is of the form $H = H_0 + \varepsilon V$, where $H_0$ is an integrable Hamiltonian and $\varepsilon V$ is a small perturbation with $\varepsilon \ll 1$, we can obtain a control term $f(\varepsilon V)$ in order to make the system more regular. From expression (5), we see that $f$ is given by an infinite sum of terms, but the literature shows us that it is possible to achieve good results even if we keep just one or two terms in $f$ [28--30], which proves the robustness of the theory. Moreover, the truncation of $f$ is convenient, and sometimes even necessary, since the control term should be as simple as possible in order to be implemented experimentally.

Using expressions (8) to (11) and keeping just the terms up to $n=1$ in expression (5), the control term for our system is given by

$$f(\varepsilon V) \cong -\frac{1}{2}\{\Gamma(\varepsilon V)\}(\varepsilon V)$$

$$f(\varepsilon V) \cong \frac{\varepsilon^2}{8T^2}\sin^2\left(k\sqrt{2I}\sin\theta\right)\left[\sum_{n=-\infty}^{+\infty}\cos\left(\frac{2\pi n t}{T}\right)\right]^2$$
$$\times \left[\frac{\sqrt{1+2I}}{2I} + \frac{\sqrt{1+2I}}{2I}\tan^2\theta - \frac{1}{\sqrt{1+2I}}\right]$$

$$f(\varepsilon V) \cong \frac{\varepsilon^2}{16}\left[1 - \cos\left(2k\sqrt{2I}\sin\theta\right)\right]\sum_{n=-\infty}^{+\infty}\delta(t-nT)$$
$$\times \left[\frac{\sqrt{1+2I}}{2I} + \frac{\sqrt{1+2I}}{2I}\tan^2\theta - \frac{1}{\sqrt{1+2I}}\right]$$

($I \neq 0$ and $\theta \neq \pi/2, 3\pi/2$ for $0\leq \theta < 2\pi$). (12)

From expression (12), we notice that the terms proportional to $(\partial H_0/\partial E)(\partial V/\partial t)$ canceled each other. However, we point out that this is a particular feature of our system. It does not happen for all Hamiltonians as can be seen, for example, in Refs. [29, 33] where these terms do not get canceled.

In order to avoid indeterminations and an unlimited growth of $f(\varepsilon V)$ in specific regions of phase space, we will truncate the control term given in (12) and we will consider only the last term of this expression. Besides, we will expand $(1+2I)^{-1/2}$ in a Taylor series centered at $I_0$ as

$$\frac{1}{\sqrt{1+2I}} = \frac{1}{\sqrt{1+2I_0}} - \frac{I-I_0}{(1+2I_0)^{3/2}}$$
$$+ \frac{3(I-I_0)^2}{2(1+2I_0)^{5/2}} - \ldots = A + O(I), \quad (13)$$

where $A$ represents all the constant terms of the expansion and $O(I)$ represents all the terms proportional to $I^m$ with $m = 1, 2, 3, \ldots$. Truncating expression (12) as mentioned before and replacing expansion (13) into it, the resulting control term will be

$$f(\varepsilon V) = \frac{\varepsilon^2}{16}\sum_{n=-\infty}^{+\infty}\delta(t-nT)\times\left[A\cos\left(2k\sqrt{2I}\sin\theta\right)\right.$$
$$\left.+ O(I)\cos\left(2k\sqrt{2I}\sin\theta\right) - A - O(I)\right]. \quad (14)$$

Looking at equation (14), we see that the first term is very similar to the perturbative term of Hamiltonian (3), having the same physical interpretation, i.e., as that of a stationary electrostatic wave. Because of this, we choose to truncate the control term $f(\varepsilon V)$ once more and work just with the first term of equation (14)

$$f(\varepsilon V) = \varepsilon^2 a\cos\left(2k\sqrt{2I}\sin\theta + \gamma\right)\sum_{n=-\infty}^{+\infty}\delta(t-nT). \quad (15)$$

As we are not working with the entire control term given by the infinite series of expression (5), we let the amplitude and the phase of the electrostatic wave be two parameters that we can vary in order to achieve the best results. Numerical simulations show us that it happens for $a = 1/8$ and $\gamma = \pi$.

Thus, the final control term that we are going to apply to our system is given by expression (16)

$$f(\varepsilon V) = \frac{\varepsilon^2}{8}\cos\left(2k\sqrt{2I}\sin\theta + \pi\right)\sum_{n=-\infty}^{+\infty}\delta(t-nT) \quad (16)$$

and $f(\varepsilon V)$ corresponds simply to a stationary electrostatic wave of wave vector $2k$, period $T$, amplitude $\varepsilon^2/8$ and phase $\pi$ lying along the $x$ axis.

## IV. ANALYTICAL AND NUMERICAL RESULTS

The Hamiltonian of the system described in Section II with the addition of the control term (16) is given by

$$H = \sqrt{1+2I} + \left[\frac{\varepsilon}{2}\cos\left(k\sqrt{2I}\sin\theta\right) + \frac{\varepsilon^2}{8}\cos\left(2k\sqrt{2I}\sin\theta + \pi\right)\right]\sum_{n=-\infty}^{+\infty}\delta(t-nT) \quad (17)$$

and following once again the procedure described in Section II it is possible to obtain an explicit map that relates the variables $I$ and $\theta$ at kicks $n$ and $n+1$

$$I_{n+1} = \frac{1}{2}\left\{2I_n\sin^2\theta_n + \left[\sqrt{2I_n}\cos\theta_n + \frac{1}{2}\varepsilon k\sin\left(k\sqrt{2I_n}\sin\theta_n\right) + \frac{1}{4}\varepsilon^2 k\sin\left(2k\sqrt{2I_n}\sin\theta_n + \pi\right)\right]^2\right\},$$

(18)

$$\theta_{n+1} = \arctan\left(\frac{4\sqrt{2I_n}\sin\theta_n}{4\sqrt{2I_n}\cos\theta_n + 2\varepsilon k\sin\left(k\sqrt{2I_n}\sin\theta_n\right) + \varepsilon^2 k\sin\left(2k\sqrt{2I_n}\sin\theta_n + \pi\right)}\right) + \frac{T}{\sqrt{1+2I_{n+1}}}.$$

Figure 1 contains the phase spaces constructed from maps (4) and (18) for $T = 2\pi(1+1/15)$, $k = 2$ and $\varepsilon = 0.2$. Panel (a) shows the system without the control term and we see that the chaotic sea fills a great part of the phase space. Only the region of low action $I$ and some islands remain regular. Panel (b) illustrates the system with the addition of the control term $f$. In contrast to panel (a), the phase space in (b) is regular in almost all its regions. Chaos can be seen in panel (b) just around some of the islands of the system because of the hyperbolic points located between the islands of a chain.

For $\varepsilon = 0.2$ as in Fig. 1, the amplitude of the control term $f$ is just 5.0% of the amplitude of $\varepsilon V$, i.e., for generating the electrostatic wave described by $f$, we spend only 5.0% of the energy used to produce the electrostatic wave $\varepsilon V$. This is one of the most remarkable features in the procedure used to determine $f(\varepsilon V)$: it allows one to control chaos in the system with little energy cost. This together with the simple form of $f$ makes it possible to implement the control term experimentally.

We also point out that the controlled system presents the same resonances as the original system, for example, the (1,1) resonance located about $I_{1,1} \approx 0.07$, the (4,3) resonance located about $I_{4,3} \approx 0.51$, the (3,2) resonance for which $I_{3,2} \approx 0.78$ and the (2,1) resonance located around $I_{2,1} \approx 1.78$ (see the Appendix for more details about the system resonances). It means that the addition of a small and suitable control term does not change the main structures of the system although it drastically reduces chaos in phase space.

In Ref. [32], we studied the regular acceleration experienced by the particle in the region of low action $I < 0.20$. However, it is also possible to coherently accelerate the particle in regions of higher action as can be seen in Fig. 1 for $1.00 < I < 2.70$. Considering the islands centered at $\theta \approx 0; \pi$, we observe that the value of the action increases when the electrostatic wave transfers energy to the particle. In panel (a), the most external trajectories of the islands centered at $I_{2,1} \approx 1.78$ and $\theta \approx 0; \pi$ have been destroyed by chaos and just the internal ones survive. In panel (b), the addition of the control term to the system recovers the most external trajectories of these islands, improving the process of regular acceleration.

Numerical calculations enable us to determine the maximum width of the islands centered at $I_{2,1} \approx 1.78$ and $\theta \approx 0; \pi$. Using this information, it is possible to calculate the minimum and maximum dimensionless energies of the particle inside the islands. For the system without the control term shown in Fig. 1(a), the minimum energy of the particle is $E_{\min} \approx 1.87$ and its maximum energy is $E_{\max} \approx 2.40$. With the addition of the control term, the minimum energy of the particle in the islands is $E_{\min} \approx 1.77$ and its maximum energy is $E_{\max} \approx 2.50$.

Then, we see that by adding the control term to the system, the minimum energy of the particle in the islands is 5.35% lower than its minimum energy in the original system. On the other hand, the maximum energy of the particle is 4.17% higher in the controlled system. It means that, in the controlled dynamics, the initial energy of the particle can be lower than in the original system and even so it will be more accelerated by the wave since its final energy is higher.

We may try to understand how the coupling of the two electrostatic waves happens and the role of $f$ in the phase space shown in Fig. 1(b) by studying a Hamiltonian where the only perturbation is the control term. Figure 2 presents the phase space of the system described by Hamiltonian (19) for the same parameters used in Fig. 1

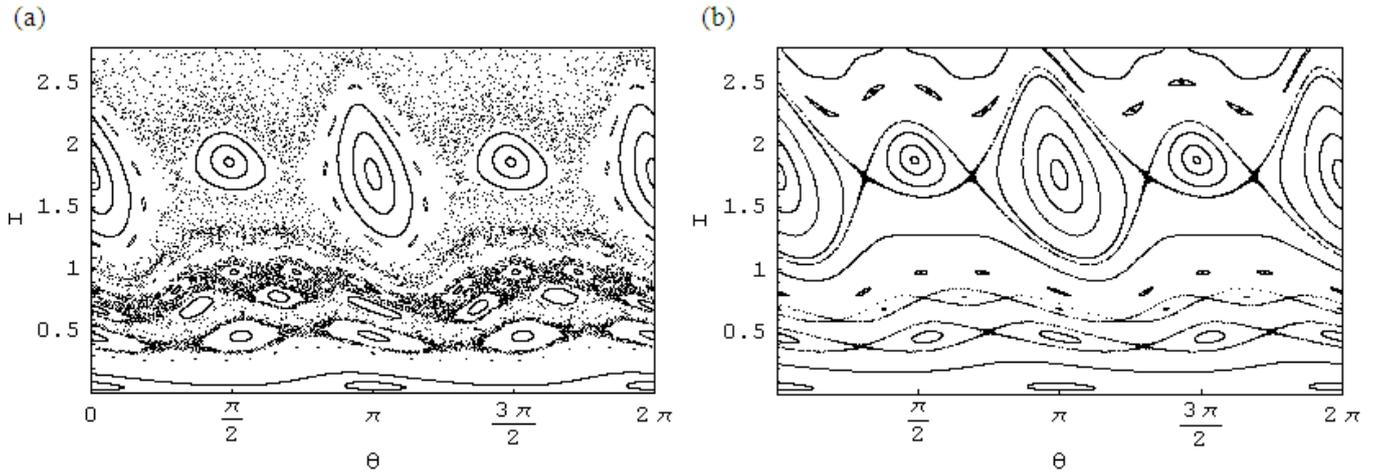

FIG. 1. Phase spaces of the system for $T = 2\pi(1+1/15)$, $k = 2$ and $\varepsilon = 0.2$. Panel (a) illustrates the system without the control term while Panel (b) shows the system with the addition of the control term.

$$H = H_0 + f = \sqrt{1+2I}$$
$$+ \frac{\varepsilon^2}{8}\cos\left(2k\sqrt{2I}\sin\theta + \pi\right)\sum_{n=-\infty}^{+\infty}\delta(t-nT). \quad (19)$$

We see that almost all the trajectories displayed in Fig. 2 are regular. Chaos is present only in very small regions around some of the islands of the system because of the hyperbolic points that exist between the islands of a chain.

One remarkable feature of the control term $f(\varepsilon V)$ obtained is that it was constructed in such a way that makes Hamiltonian (19) present exactly the same resonances as the original system described by Hamiltonian (3) (see the Appendix for more details about the system resonances). For example, the resonances (1,1), (2,1), (3,2) and (4,3) are located in the same position in $I$ in both Figs. 1(a) and 2. However, the position of the islands with respect to the $\theta$ variable is different in the two phase spaces, as well as the stability of some of the equilibrium points and the number of islands present in each resonance.

Comparing all the phase spaces illustrated in this work, we see that the system perturbed by the two electrostatic waves $\varepsilon V + f$ presents a behavior that is a mixture of the individual behaviors of the systems described by Hamiltonians (3) and (19). The amplitude of $f$ is much smaller than the amplitude of $\varepsilon V$ as well as the size of the islands created by $f$. Thus, the structure of islands in Fig. 1(b) is the same as the one in Fig. 1(a) that corresponds to the system without the control term.

Looking now to the region outside the main islands of the system, we see that almost all this region is filled by chaos in Fig. 1(a), while in Fig. 2 all the trajectories are regular. In Fig. 1(b), the region outside the islands is regularized, presenting invariant tori. From this, we can conclude that even being much smaller than $\varepsilon V$, the behavior associated with the control term $f$ is predominant in the region we are considering. The addition of $f$ to the system creates invariant tori in the whole phase space preventing chaos from occurring and allowing the islands of the (2,1) resonance to grow more than in the original system.

## V. CONCLUSIONS

We analyzed the interaction of a magnetized relativistic particle with a stationary electrostatic wave given as a series of pulses. From the Hamiltonian of the system, we obtained a fully explicit map that describes its time evolution. The map allowed us to build the phase space of the system and then to analyze whether its behavior is regular or chaotic.

We also applied the method of control developed in Ref. [27] for near-integrable Hamiltonians to calculate a control term for our system. In this case, the control term obtained gave us simply a second stationary electrostatic wave that should be added to the system. In addition, the amplitude of our control term is very small compared to the perturbation originally applied to the system. Such a small control term does not alter the structure of islands of the system, but the controlled dynamics is much more regular. While in the original system chaos fills a great area of the phase space, in the controlled system, chaos is limited to a small region and almost all the trajectories are regular.

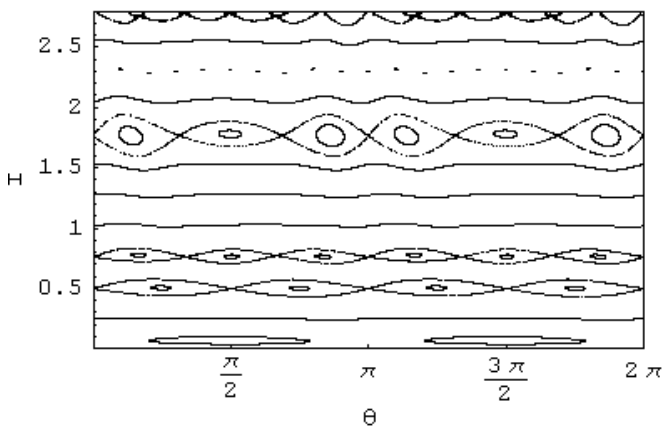

FIG. 2. Phase space of the system perturbed only by the control term. The figure was built using the parameters $T = 2\pi(1+1/15)$, $k = 2$ and $\varepsilon = 0.2$.

Besides regularizing the system, we showed that the control of chaos can be used to improve regular particle acceleration. In the controlled dynamics, the suppression of chaos restores periodic and quasiperiodic trajectories that are responsible for coherently accelerating the particle. Therefore, in the controlled system, a particle with initial energy lower than in the original system can be more accelerated and can achieve a final energy higher than it would in the system without control.

Finally, we built a phase space considering just the interaction of the magnetized relativistic particle with the control term. It allowed us to present a qualitative analysis about how the control term acts on the original system and how the coupling of the two electrostatic waves that gives rise to the features of the controlled dynamics happened.

## ACKNOWLEDGMENTS

We acknowledge financial support from the Brazilian scientific agencies FAPESP and CNPq.

## APPENDIX: RESONANCES OF THE SYSTEM

To calculate the primary resonances of the original system, it is necessary to write the periodic collection of delta functions of Hamiltonian (3) as a Fourier series

$$H = \sqrt{1+2I} + \frac{\varepsilon}{2T}\cos\left(k\sqrt{2I}\sin\theta\right)\sum_{s=-\infty}^{+\infty}\cos\left(\frac{2\pi s t}{T}\right)$$

$$H = \sqrt{1+2I} + \frac{\varepsilon}{4T}\sum_{s=-\infty}^{+\infty}\left[\cos\left(k\sqrt{2I}\sin\theta+\frac{2\pi s t}{T}\right)\right.$$
$$\left. + \cos\left(k\sqrt{2I}\sin\theta-\frac{2\pi s t}{T}\right)\right]. \quad (A1)$$

Besides, we will expand the cosine functions of Hamiltonian (A1) in a Fourier-Bessel series using relation (A2)

$$\cos\left(k\sqrt{2I}\sin\theta\pm\frac{2\pi s t}{T}\right)$$
$$= \sum_{r=-\infty}^{+\infty}J_r\left(k\sqrt{2I}\right)\cos\left(r\theta\pm\frac{2\pi s t}{T}\right) \quad (A2)$$

where $J_r\left(k\sqrt{2I}\right)$ are Bessel functions of the first kind. Replacing relation (A2) into (A1), the Hamiltonian of the system will be given by

$$H = \sqrt{1+2I} + \frac{\varepsilon}{4T}\sum_{s=-\infty}^{+\infty}\sum_{r=-\infty}^{+\infty}\left\{J_r\left(k\sqrt{2I}\right)\right.$$
$$\left.\times\left[\cos\left(r\theta+\frac{2\pi s t}{T}\right)+\cos\left(r\theta-\frac{2\pi s t}{T}\right)\right]\right\}. \quad (A3)$$

From Hamiltonian (A3), we determine the primary resonances of the system as

$$\frac{d}{dt}\left(r\theta\pm\frac{2\pi s t}{T}\right) = 0, \quad r\frac{d\theta}{dt} = \mp\frac{2\pi s}{T}$$
$$r\omega_0 \cong \mp s\omega \quad (A4)$$

where $\omega = 2\pi/T$ is the frequency of the electrostatic wave and we approximated $d\theta/dt$ as the natural frequency $\omega_0$ of the unperturbed motion,

$$\omega_0 = \left.\frac{d\theta}{dt}\right|_{H=H_0} = \frac{dH_0}{dI} = \frac{1}{\sqrt{1+2I}}. \quad (A5)$$

Replacing expression (A5) into (A4), we determine the values of the action $I$ for which the system is resonant

$$I_{r,s} \cong \frac{1}{2}\left(\frac{r}{s\omega}\right)^2 - \frac{1}{2} = \frac{1}{8}\left(\frac{rT}{s\pi}\right)^2 - \frac{1}{2}. \quad (A6)$$

From expression (A6), we see that the position of the $(r, s)$ resonances in phase space depends on the period of the electrostatic wave $\varepsilon V$ and on the ratio between two integers $r \geq 1$ and $s \geq 1$.

Numerical simulations tell us that $r$ represents the number of islands in a chain while $s$ is proportional to the change in the $\theta$ variable between two consecutive kicks, i.e., $\Delta\theta = \theta_{n+1} - \theta_n \cong 2\pi s/r$ (mod $2\pi$).

For some $(r, s)$ resonances, the system presents only one chain of $r$ islands. This is the case of the (4,3) resonance for which $I_{4,3} \approx 0.51$. The central elliptic points of each island correspond to a single periodic orbit. The trajectory moves from one point to the other and between two consecutive kicks, we have $\Delta\theta \cong 2\pi s/r = 3\pi/2$ (mod $2\pi$).

However, for some $(r, s)$ resonances, we find not one but two chains of $r$ islands. It happens for the (1,1), (2,1) and (3,2) resonances shown in Fig. 1. For example, the (2,1) resonance located around $I_{2,1} \approx 1.78$ presents two chains with two islands each. The central elliptic points at $\theta \approx 0; \pi$ and $\theta \approx \pi/2; 3\pi/2$ form two distinct periodic orbits. For both chains, the trajectory of the particle moves from one island to the other repeatedly, such that, between two consecutive kicks, we have $\Delta\theta \cong 2\pi s/r = \pi$ (mod $2\pi$).

We can also determine the primary resonances of Hamiltonians (17) and (19). Following the same procedure described in this Appendix, we see that the position of the resonances in phase space is also given by expression (A6). Both the system without the control term and the controlled system present primary resonances for the values of action $I_{r,s}$ given in (A6).

Besides, the resonances of Hamiltonians (3) and (17) are at the same position in phase space with respect to the $\theta$ variable as can be seen in Fig. 1. However, the position with respect to $\theta$ is different for the resonances of Hamiltonian (19) as well as the stability of some of the equilibrium points and the number of chains present in each resonance as can be seen comparing Figs. 1 and 2.